\documentclass[twocolumn, showpacs, preprintnumbers,amsmath,amssymb,balancelastpage]{revtex4}
\usepackage{putexPRL}
\usepackage{graphicx}
\usepackage{dsfont}
\usepackage{amsmath}
\usepackage{array}
\usepackage{epstopdf}
\usepackage{enumerate}
\usepackage{youngtab}
\usepackage{tensor}
\usepackage{braket}
\usepackage[normalem]{ulem}
\usepackage{slashed}
\usepackage[aligntableaux=center]{ytableau}
\usepackage[utf8]{inputenc}
\usepackage[
      colorlinks=true,
      linkcolor=blue,
      urlcolor=blue,
      filecolor=black,
      citecolor=red,
      ]{hyperref}
\usepackage{braket}
\usepackage{mathtools}
\usepackage{rotating}
\usepackage{tabularx}
\newcolumntype{Y}{>{\centering\arraybackslash}X}
\newcolumntype{C}[1]{>{\centering\arraybackslash}p{#1}}
\usepackage{xcolor}
\definecolor{LightCyan}{rgb}{0.7,1,1}
\definecolor{Gray}{gray}{0.9}
\usepackage{xspace}

\newcommand{\grp}[1]{\mathrm{#1}}

\newcommand{\grU}{\grp{U}}
\newcommand{\grSU}{\grp{SU}}

\newcommand{\abs}[1]{\left\lvert #1 \right\rvert}

\newcommand {\be} {\begin {equation}}
\newcommand {\ee} {\end {equation}}

\newcommand {\bes} {\begin {equation*}}
\newcommand {\ees} {\end {equation*}}

\newcommand{\es}[2] {\begin{equation} \label{#1} \begin{split} #2 \end{split} \end{equation}}

\newcommand{\beq}{\begin{equation}}
\newcommand{\eeq}{\end{equation}}

\def\ie{\begin{equation}\begin{aligned}}
\def\fe{\end{aligned}\end{equation}}

\def\<{\langle}
\def\>{\rangle}

\def\beg{\begin{equation}\begin{gathered}}
\def\eeg{\end{gathered}\end{equation}}

\def\bea{\begin{equation}\begin{aligned}}
\def\eea{\end{aligned}\end{equation}}

\newcommand{\mlata}{m_{\text{lat},\alpha}}

\begin{document} 

\preprint{PUPT-2640}

\title{Phase Diagram of the Two-Flavor Schwinger Model at Zero Temperature}

\author{{Ross Dempsey,$^1$ Igor R.~Klebanov,$^{1, 2}$ Silviu S.~Pufu,$^{1,2}$ Benjamin T. S\o gaard,$^1$ and Bernardo Zan$^1$}}         

\affiliation{$^{1}$Joseph Henry Laboratories, Princeton University, Princeton, NJ 08544, USA}
\affiliation{$^{2}$Princeton Center for Theoretical Science, Princeton University, Princeton, NJ 08544, USA}

\begin{abstract}
We examine the phase structure of the two-flavor Schwinger model as a function of the $\theta$-angle and the two masses, $m_1$ and $m_2$. 
In particular, we find interesting effects at $\theta=\pi$:  along the $\grSU(2)$-invariant line $m_1 = m_2 = m$, in the regime where $m$ is much smaller than the charge $g$, 
the theory undergoes logarithmic RG flow of the Berezinskii-Kosterlitz-Thouless type. As a result, dimensional transmutation takes place, leading to a non-perturbatively small mass gap $\sim e^{- A g^2/m^2}$. The $\grSU(2)$-invariant line lies within a region of the phase diagram where the charge conjugation symmetry is spontaneously broken and whose boundaries we determine numerically. Our numerical results are obtained using the Hamiltonian lattice gauge formulation that includes the mass shift $m_{\rm lat}= m- g^2 a/4$ dictated by the discrete chiral symmetry.    
\end{abstract}


\maketitle
\nopagebreak

\section{Introduction}

Quantum Electrodynamics (QED) in $1+1$ dimensions, also known as the Schwinger model \cite{Schwinger:1962tp}, is a famous model of Quantum Field Theory (QFT) that has played an important role for over 60 years \cite{Lowenstein:1971fc,Casher:1974vf, Coleman:1975pw}. It is a useful theoretical laboratory for various important phenomena, including QFT anomalies and confinement of charge. Its lattice Hamiltonian implementations~\cite{Kogut:1974ag,Banks:1975gq} have connections with condensed matter and atomic physics, and, in recent years, there have been efforts to construct experimental setups for its quantum simulations (for a review, see \cite{Banuls:2019bmf}).
 
The model with one massless Dirac fermion of charge $g$ is exactly solvable, reducing to the non-interacting Schwinger boson of mass $M_S= g/\sqrt \pi$; this can be concisely demonstrated via the bosonization of the fermion~\cite{Coleman:1975pw}. The $\grU(1)$ chiral symmetry of the massless action is broken by the Schwinger anomaly. The massive model, in addition to containing the obvious dimensionless parameter $m/g$, depends on the $\theta$ angle related to the introduction of a background electric field \cite{Coleman:1975pw}. This parameter, which has periodicity $2 \pi$, is somewhat analogous to the $\theta$ angle of the $3+1$ dimensional gauge theory.

Generalizations of the Schwinger model to $N_f>1$ flavors of fermions of charge $g$ exhibit a richer set of phenomena \cite{Coleman:1976uz}. When the fermions are massless, 
the Schwinger model has $\grSU(N_f)\times \grSU(N_f)$ chiral symmetry. Its low-energy limit is described \cite{Gepner:1984au,Affleck:1985wa} by the $\grSU(N_f)_1$ Wess-Zumino-Witten (WZW) model, which is a Conformal Field Theory (CFT) of central charge $N_f-1$. The $N_f > 1$ Schwinger model also contains a massive sector that includes the Schwinger boson. Therefore, it was hoped that the multiflavor Schwinger models may provide simple realizations of the ``unparticle physics" idea \cite{Georgi:2007ek}, and this motivated the papers \cite{Georgi:2019tch,Georgi:2020jik,Georgi:2022sdu}. As in these papers, we will focus on $N_f = 2$, where for $m=0$ the IR CFT is described by a compact scalar at the self-dual radius. While investigations of this model have a long history including \cite{Coleman:1976uz,Steinhardt:1977tx,Smilga:1992hx,Hetrick:1995wq,Smilga:1998dh,Hosotani:1998kd,Berruto:1999ga,Hip:2021jgp,Albergo:2022qfi,Funcke:2023lli}, we will present a number of new results:
1) Even in the limit of small masses, we can have spontaneous symmetry breaking of the charge conjugation symmetry, or critical behavior, or an IR trivial phase.  2) For $\theta=\pi$ and $m/g\ll 1$, there is an effective field theory description in terms of the sine-Gordon model with $\beta\approx \sqrt{8\pi}$ \cite{Coleman:1976uz,Smilga:1998dh}.
We describe the $\grSU(2)$-invariant RG trajectory, which flows from asymptotic freedom in the UV, and in the IR it produces an exponentially small mass gap $\sim e^{- A g^2/m^2}$, with $A \approx 0.111$ as we show below.
Therefore, the $N_f = 2$ Schwinger model with  $\theta=\pi$ has some qualitative similarities with QCD because it can exhibit {\it dimensional transmutation}. 

\begin{figure}[htbp]
\begin{center}
 \includegraphics[width=0.3\textwidth]{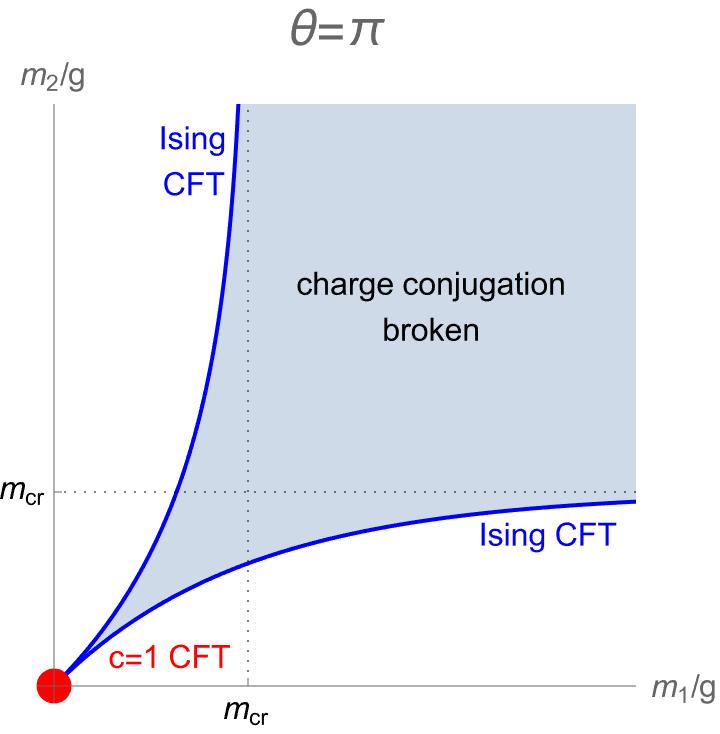}
\caption{The schematic phase diagram for the 2-flavor Schwinger model at $\theta = \pi$ (it is similar to the phase diagram of the 2-flavor QCD, exhibited in
\cite{Creutz:2010ts,Creutz:2018vgl}).
In the shaded region, charge conjugation symmetry is broken, and there are two degenerate vacua.  This region is bounded by two critical curves of Ising CFTs.  These two curves meet at the origin, where the low-energy description is provided by the $\grSU(2)_1$ WZW model. For $m_1=m_2\ll g$, the model exhibits an exponentially large correlation length due to dimensional transmutation. 
}
\label{PDFigure}
\end{center}
\end{figure}

We discuss the zero-temperature phase diagram as a function of $\theta$ and the masses $m_1$ and $m_2$ of the two fermion flavors, which we can restrict to be positive (some aspects of the phase structure were discussed in the past \cite{Coleman:1976uz,Smilga:1998dh,Georgi:2022sdu}).  Our proposal is that, while for all $\theta \neq \pi$ this model has a non-degenerate vacuum, for $\theta = \pi$ the phase diagram is as in Figure~\ref{PDFigure}. It contains two critical curves that pass through the origin, along which the low-energy physics is governed by the 2D Ising CFT of central charge $c=1/2$.  In the shaded region of Figure~\ref{PDFigure}, the charge conjugation symmetry $C$, defined below, is spontaneously broken, leading to two degenerate vacua. This phenomenon, which was recently studied in \cite{Funcke:2023lli}, is reminiscent of the spontaneous breaking of $CP$ symmetry in 4D Yang-Mills theory at $\theta=\pi$ \cite{Dashen:1970et,Creutz:1995wf,Smilga:1998dh,Creutz:2010ts,Creutz:2018vgl,Gaiotto:2017yup}, and there are analogous phenomena in 2D scalar QED \cite{Komargodski:2017dmc,Sulejmanpasic:2020lyq}.   
We present both analytical and numerical evidence for the phase diagram in Figure~\ref{PDFigure}.  On the numerical side, our calculations using the Hamiltonian lattice approach are in excellent agreement with the continuum analysis.  The convergence of the numerical calculations is significantly improved by including the mass shift (\ref{eq:mass_shift}) derived in \cite{Dempsey:2022nys}.

\section{The setup}

Let us consider the Schwinger model with $N_f$ fermion flavors of masses $m_\alpha$, with $\alpha = 1, \ldots, N_f$.  While the $m_\alpha$ are, in general, complex parameters, the $\grU(1)^{N_f}$ axial transformations can be used to set all $m_\alpha$ real with $m_\alpha \geq 0$. Then, the Lagrangian density is 
 \es{LagDens}{
  {\cal L} &=  - \frac{1 }{4g^2}F_{\mu\nu}^2  - \frac{\theta}{4 \pi} \epsilon^{\mu\nu} F_{\mu\nu} 
     +\sum_{\alpha = 1}^{N_f} \overline \Psi_\alpha (i \slashed{D}  - m_\alpha) \Psi_\alpha .    
 }
Here, $\epsilon^{01}=1$, $\slashed{D} = \gamma^\mu (\partial_\mu + i A_\mu)$, and $(\gamma^0, \gamma^1) = (\sigma_3, i \sigma_2)$ obey $\{\gamma^\mu, \gamma^\nu\} = 2 \eta^{\mu\nu} = 2 \diag\{1, -1\}$.  

To study this model numerically, we use the Hamiltonian lattice formulation of \cite{Kogut:1974ag,Banks:1975gq}, where the spatial direction is discretized into $N$ sites, with $N$ even, while the time direction remains continuous. The two-component Dirac fermions of each flavor are staggered, with the $\gamma^0$ eigenstates of eigenvalue $+1$ and $-1$ being placed on even and odd sites, respectively.  The lattice Hamiltonian is
 \es{eq:gauge_hamiltonian}{
    H &= \frac{g^2 a}{2} \sum_{n=0}^{N-1} \left(L_n + \frac{\theta}{2\pi}\right)^2 + \sum_{\alpha = 1}^{N_f} \mlata  \sum_{n=0}^{N-1}(-1)^n c^\dagger_{n,\alpha} c_{n,\alpha}\\
    &{} -\frac{i}{2a} \sum_{n=0}^{N-1} \sum_{\alpha = 1}^{N_f} \left(c^\dagger_{n,\alpha} U_n c_{n+1,\alpha}  - c^\dagger_{n+1,\alpha} U_n^\dagger c_{n,\alpha}\right) \,.
 }  
Here, $a$ is the lattice spacing, $c_{n,\alpha}$  and $c_{n,\alpha}^\dagger$ are the annihilation and creation operators for a fermion of flavor $\alpha$ on site $n$, and $U_n = e^{i \phi_n}$ is a unitary operator living on the link between sites $n$ and $n+1$. The electric field strengths $L_n = -i \frac{\partial}{\partial \phi_n}$ are integer-valued, while $\theta \in [0,2\pi)$ comes from the $\theta$-term in the action, and $\frac{\theta}{2 \pi}$ acts as a fractional background electric field.  The Hamiltonian should be supplemented by the Gauss law constraint
 \es{Gauss}{
  L_n - L_{n-1} &= \sum_\alpha  
   \left( c_{n, \alpha}^\dagger c_{n, \alpha} - \frac{1 - (-1)^n}{2} \right) \,.
 }

The parameters $g$ and $\theta$ of the lattice model should be identified with the analogous  parameters of the continuum model \eqref{LagDens}.  As argued in~\cite{Dempsey:2022nys}, one should take
 \es{eq:mass_shift}{
    \mlata = m_\alpha - \frac{N_f g^2 a}{8} \,.
 }
In \cite{Dempsey:2022nys}, it was also shown that when $N_f$ is even and $m_\alpha = 0$, the lattice theory is invariant under translation by one site, which corresponds to a discrete chiral symmetry in the continuum. In the leading strong coupling limit, where the hopping term is ignored, the ground state can be highly degenerate. For $N_f=2$ and $m_1=m_2=0$, we find that the strong coupling degeneracy is $3^N+1$ for $\theta=0$, while it is $2^N$ for $\theta =\pi$. The latter fact provides a starting point for the correspondence between the $N_f=2$ Schwinger model at $\theta=\pi$ and the Heisenberg antiferromagnet \cite{Hosotani:1998kd,Berruto:1999ga}.  

The integrated fermion bilinear operator $\int dx\,  \bar \Psi \Psi$ translates into the lattice operator 
$\sum_{n=0}^{N-1}(-1)^n c^\dagger_{n,\alpha} c_{n,\alpha}$,
which is odd under the unit shift. The uniqueness of the ground state away from the strong coupling limit for $m_\alpha = 0$, and the symmetry under the unit translation, imply that the VEV of the mass operator vanishes on a periodic lattice with an even number of sites.

When $\theta = 0$ or $\pi$, for any $m_\alpha$, the models \eqref{LagDens} and \eqref{eq:gauge_hamiltonian}  are invariant under a charge conjugation symmetry $C$.  In the continuum, $C$ acts as 
 \es{CCont}{
   \text{$C$}: \qquad  A_\mu \to - A_\mu \,, \qquad \Psi_\alpha \to \gamma^5 \Psi_\alpha^* \,,
 }
where $\gamma^5 = \gamma^0 \gamma^1 = \sigma_1$, and on the lattice it acts as \cite{Berruto:1999ga}
 \es{CLattice}{
  \text{$C$}: \qquad  L_n &\to - L_{n+1} - \frac{\theta}{\pi} \,, \qquad U_n \to U_{n+1}^\dagger \,, \\
   c_{n, \alpha} &\to c_{n+1, \alpha}^\dagger \,, \qquad
     c_{n, \alpha}^\dagger \to c_{n+1, \alpha} \,.
 }
It is this symmetry $C$ that is broken whenever there are two degenerate vacua in the phase diagram in Figure~\ref{PDFigure}.

For any $\theta$ and $m_\alpha = m$ for all $\alpha$, the models \eqref{LagDens} and \eqref{eq:gauge_hamiltonian} are invariant under an $\grSU(N_f)$ symmetry under which the fermions transform in the fundamental representation.  When $m=0$, in the continuum model \eqref{LagDens}, this $\grSU(N_f)$ symmetry is enhanced to $\grSU(N_f) \times \grSU(N_f)$.

\section{Continuum Treatment of the Two-flavor Schwinger model }

Let us now take $N_f = 2$ and present continuum field theory arguments in support of the phase diagram of the 2-flavor Schwinger model in Figure~\ref{PDFigure}. 

\subsection{One very massive fermion}

When one of the fermions is very massive, it can be integrated out, leaving us with the $N_f = 1$ model.  For a fermion mass $m$, the phase diagram of the $N_f = 1$ model exhibits a line of first order phase transitions at $\theta = \pi$ that extends over the interval $(m_\text{cr}, \infty)$ \cite{Coleman:1976uz} , with $m_\text{cr} \approx 0.33 g$.   At $m = m_\text{cr}$, there is evidence \cite{Byrnes:2002gj,Byrnes:2002nv} that the second order phase transition is in the 2D Ising universality class.  For $m > m_\text{cr}$ and $\theta = \pi$, there are two degenerate vacua, each of which breaks $C$ spontaneously.  Everywhere else on the phase diagram there is a non-degenerate vacuum and a non-zero gap.

Without loss of generality, suppose we take $m_2 / g \gg 1$.  Integrating out $\Psi_2$ in \eqref{LagDens} yields the effective Lagrangian
 \es{EffLag}{
  {\cal L}^{N_f = 1} + \frac{1}{2} \int d^2 y\, A_\mu(x) A_\nu(y) \Pi_2^{\mu\nu}(x-y) + O(A^4) \,,
 }
where $\Pi_2^{\mu\nu}  = (- \partial^2 \eta^{\mu\nu} + \partial^\mu \partial^\nu) \Pi_2(x)$ is the one-loop vacuum polarization.  The Fourier transform $\Pi_2(q) = \int d^2x\,  \Pi_2(x) e^{i q\cdot x} $ is (see (7.90) of \cite{Peskin:1995ev}) \footnote{The result in (7.90) of \cite{Peskin:1995ev} should be divided by a factor of $2e^2$.  The division by $e^2$ is due to our normalization of the gauge kinetic term, and the division by $2$ is due to the fact that for two-component spinors $\tr {\bf 1} = 2$ as opposed to $\tr {\bf 1} = 4$ as was assumed in \cite{Peskin:1995ev}.}
 \es{Pid2}{
  \Pi_2(q) = - \frac{1}{\pi} \int_0^1 d\xi\, \frac{\xi(1-\xi)}{m_2^2 - \xi(1-\xi) q^2} \approx - \frac{1}{6 \pi m_2^2} 
 }
at large $m_2$.  Thus, $\Pi_2(x) \approx  - \frac{1}{6 \pi m_2^2} \delta^{(2)}(x)$, and the effective Lagrangian \eqref{EffLag} 
becomes, approximately, that of the one-flavor model with an effective gauge coupling:
 \es{EffectiveLargeMass2}{
  g^{-2}_\text{eff}= g^{-2}  \left( 1 + \frac{1}{6 \pi} \frac{g^2}{m_2^2} 
   \right) \,.
 }
Since the one-flavor Schwinger model exhibits an Ising second-order phase transition at $m_{\text{cr}} \approx 0.33 g$ at $\theta = \pi$, it follows that the 2-flavor Schwinger model with $m_2 / g \gg 1$ also exhibits an Ising phase transition at $\theta = \pi$ for $m_{\text{cr}} \approx 0.33 g_\text{eff}$. Expanding this we get
 \es{m1crit}{
  m_{1, \text{cr}}(m_2) \approx 0.33 g \left( 1 - \frac{1}{12 \pi} \frac{g^2}{m_2^2} +  O\left ( \frac{g^4}{m_2^4} \right ) \right)\,.  
 }
The phase diagram should of course be invariant under interchanging $m_1 \leftrightarrow m_2$ so, at $\theta = \pi$, there should also be an Ising  transition at $m_{2, \text{cr}}(m_1) $ given by the RHS of \eqref{m1crit} with $m_2 \to m_1$.  The expression \eqref{m1crit} and the one obtained after interchanging $m_1 \leftrightarrow m_2$ represent the asymptotic behaviors of the blue curves in Figure~\ref{PDFigure}.  The large mass analysis also shows that in the wedge between the two curves we expect two degenerate ground states, while outside of this wedge we expect a non-degenerate ground state, just as in the $N_f=1$ model at $\theta = \pi$.  

This argument also shows that when $\theta \neq \pi$ and one of the fermions is very massive, the ground state is non-degenerate because this is also the case in the one-flavor model.  In fact, for $\theta \neq 0, \pi$ we must have a non-degenerate ground state because there is no charge conjugation symmetry that can be spontaneously broken.

\subsection{Small mass regime}

Near $m_1 = m_2 = 0$, a useful equivalent description is obtained using Abelian bosonization \cite{Coleman:1976uz}.  (One can also use non-Abelian bosonization, as in \cite{Gepner:1984au}.)  Following \cite{Coleman:1976uz}, we bosonize the fermions $\Psi_{1, 2}$ to scalar fields $\phi_{1, 2}$, and reparameterize them via $\phi_+ = 2^{-1/2}(\phi_1+\phi_2+{\textstyle\frac{1}{2}}\pi^{-1/2}\theta)$ and $\phi_- = 2^{-1/2}(\phi_1 - \phi_2)$.

Let us restrict our attention to $m_1 = m_2 = m $.  The bosonized Lagrangian is
 \es{BosLag}{
  {\cal L}_\text{bos} &= -\frac{1}{4 g^2} F_{\mu\nu}^2  
      - \frac{\phi_+}{ \sqrt{2\pi}} \epsilon^{\mu\nu} F_{\mu\nu}
       + \frac{1}{2} (\partial_\mu \phi_+)^2 
        + \frac{1}{2} (\partial_\mu \phi_-)^2 \\
        &\hspace{-0.3in}{}+ \frac{e^{\gamma}}{\pi} m \sqrt{\mu_+ \mu_-} N_{\mu_+} \cos \left[ \sqrt{2 \pi} \phi_+ - \frac{\theta}{2} \right]
         N_{\mu_-} \cos \left[ \sqrt{2 \pi} \phi_- \right] \,,
 }
where $N_{\cal M}$ means that the expression that follows is normal ordered by subtracting the two-point functions of a scalar field of mass ${\cal M}$.  A convenient choice is $\mu_+=\mu$, where $\mu$ is defined below, and $\mu_- / g \to 0$. 

For $m=0$, integrating out the gauge field shows that $\phi_+$ has mass $\mu=\sqrt{\frac{2}{\pi}} g$, while $\phi_-$ remains massless.  The field $\phi_-$ obeys the identification $\phi_- \sim \phi_- + \sqrt{2 \pi}$, which corresponds to the self-dual radius of the compact scalar.   Thus, for $m=0$ we have a massive sector described by $\phi_+$, and a sector consisting of the $c=1$ self-dual scalar CFT, which has $\grSU(2)\times \grSU(2)$ symmetry\@.  At low energies, the massive sector can also be integrated out, and we are left with the self-dual scalar CFT\@.

After integrating out the gauge field, we can integrate out $\phi_+$ order by order in $m$:
 \es{BosLag2}{
  {\cal L}_\text{bos} &=
         \frac{1}{2} (\partial_\mu \phi_-)^2 
         + m \sqrt{\mu \mu_-} \frac{e^{\gamma}}{\pi}  \left\langle {\cal O}_+(x) \right\rangle
         {\cal O}_-(x) \\
      &\hspace{-0.3in}{}+i \frac{e^{2 \gamma}}{2 \pi^2} m^2 \mu \mu_- \int d^2 y\, \left\langle {\cal O}_+(x) {\cal O}_+(y)  \right\rangle 
        {\cal O}_-(x) {\cal O}_-(y) \\
        &\hspace{-0.3in}{}+O(m^3) \,,
 }
with ${\cal O}_+\equiv N_\mu \cos \left[ \sqrt{2 \pi} \phi_+ - \frac{\theta}{2} \right]$, ${\cal O}_- \equiv  N_{\mu_-} \cos \left[ \sqrt{2 \pi} \phi_- \right]$, and  the expectation values taken in the theory of a free massive scalar field $\phi_+$ of mass $\mu$.
For $\theta \neq \pi$, $\left \langle {\cal O}_+ \right \rangle = \cos \frac{\theta}{2}$ gives the effective theory 
 \es{OpNonZeroTheta}{
   {\cal L}_\text{bos} &=  \frac{1}{2}(\partial_\mu \phi_-)^2 + m \sqrt{\mu \mu_-} \frac{e^{\gamma}}{\pi} \cos  \frac{\theta}{2} 
         N_{\mu_-} \cos \left[ \sqrt{2 \pi} \phi_- \right]  \\
        &{}+O(m^2) \,.
 } 
This is the self-dual scalar CFT deformed by the operator $\cos \left[ \sqrt{2 \pi} \phi_- \right]$ of dimension $1/2$, which triggers a RG flow to a gapped phase.  One can then show using RG scaling arguments or re-normal ordering \cite{Coleman:1976uz} that the mass gap is of order 
$\sim |m\cos (\theta/2)|^{2/3} g^{1/3}$.

When $\theta = \pi$, the coefficient of the relevant operator in \eqref{OpNonZeroTheta} vanishes.  Nevertheless, the mass deformation is not exactly marginal, because the only marginal deformation of the self-dual compact scalar CFT is the change in radius of the scalar, which breaks the symmetry to $\grU(1)\times \grU(1)$. This would be in contradiction with the $\grSU(2)$ symmetry of the equal-mass Schwinger model.  We can evaluate the $O(m^2)$ term in \eqref{BosLag2} using the propagator $G_M (x) = \langle \phi(x) \phi(0) \rangle = \frac{1}{2 \pi} K_0 ( M \sqrt{-x^2})$ of a free scalar field $\phi$ of mass $M$, which implies that $\langle {\cal O}_+(x) {\cal O}_+(y) \rangle = \sinh \left[2 \pi G_\mu (x - y) \right]$.  We also have 
 \es{OO}{
  &{\cal O}_-(x) {\cal O}_-(y) \\
  &{} =  \frac{e^{-2 \pi G_{\mu_-}(x-y)}  }{2} N_{\mu_-} \cos \left[ \sqrt{2 \pi} (\phi_-(x) + \phi_-(y)) \right]  \\
  &{}+  \frac{e^{2 \pi G_{\mu_-} (x-y)}}{2} N_{\mu_-} \cos \left[ \sqrt{2 \pi} (\phi_-(x) - \phi_-(y)) \right]  \,. 
 }
Plugging these results into \eqref{BosLag2}, changing variables to $z = \mu (y-x)$ and passing to Euclidean signature, we see that the integral receives contributions only from small $\abs{z}$.  Expanding in $\abs{z}$ and evaluating the integral gives  
 \es{eq:sine-gordon}{
	\mathcal L_\text{bos} &= \frac{1}{2}(\partial_\mu \phi_-)^2 
	+\frac{e^{3\gamma}I_sm^2}{8\pi^2 \mu^2} \biggl(2\pi e^{-2\gamma}(\partial_\mu \phi_-)^2 \\
	 &{}+ \mu_-^2 N_{\mu_-} \cos(\sqrt{8\pi}\phi_-) \biggr) + O(m^4) \,,
 }
where $I_s = 2 \pi \int_0^\infty d\xi\, \xi^2 \sinh K_0(\xi) \approx 10.08$ (see also \cite{Smilga:1998dh}). The Lagrangian \eqref{eq:sine-gordon} is that of the sine-Gordon model, a two-dimensional boson with interaction term $\sim\cos (\beta\phi)$ with $\beta>0$. By rescaling the boson to have canonical normalization, we have $\beta^2< 8\pi$.
For $m\ll g$, $\beta^2 \rightarrow 8\pi$, and the scaling dimension of the cosine operator approaches $2$. In this limit, the model is closely related to the continuum description of the Heisenberg antiferromagnet \cite{Cheng:2022sgb}.

The RG flow of the sine-Gordon model near $\beta^2=8\pi$ was computed in \cite{Kosterlitz:1974sm,Amit:1979ab} and shown to describe the BKT transition. Generically, both the coefficient of the cosine and radius of the scalar will flow.  Up to first order in the bare parameters $\alpha$ and $\delta = \frac{\beta^2}{8 \pi} - 1$, the sine-Gordon model is defined by \cite{Amit:1979ab}
 \es{LSG}{
  {\cal L} = \frac{1 - \delta}{2}(\partial_\mu \phi)^2  + \frac{\alpha e^{2 \gamma}}{32 \pi } \mu_-^2 N_{\mu_-} 
\cos (\sqrt{8 \pi} \phi) \, .
 }  
The one-loop beta functions for the running couplings $\overline \alpha$ and $\overline \delta$ are \cite{Kosterlitz:1974sm,Amit:1979ab}
 \es{Beta}{
  \beta_{\overline \alpha} = 2 \overline \alpha \overline \delta \,, \qquad
   \beta_{\overline \delta} = \frac{1}{32}  \overline \alpha^2 \,.
 }

The effective theory \eqref{eq:sine-gordon} may be restricted to have the $\grSU(2)$ symmetry that arises from the $\grSU(2)$ symmetry of the Schwinger model with equal fermion masses.   Then, in the two-dimensional parameter space $(\overline \alpha, \overline \delta)$, only the $\grSU(2)$-invariant RG trajectory can be accessed. This trajectory is the line $\overline \alpha = -8\overline \delta$ that passes through the origin, as can be seen from the fact that \eqref{eq:sine-gordon} and \eqref{LSG} imply
 \es{Gotalphadelta}{
  \alpha = \frac{8 e^\gamma I_s}{4} \frac{m^2}{g^2} = - 8 \delta \,,
 }
or from analyzing the $\grSU(2)$-invariant operators in the model \eqref{LSG}. On this locus with $\grSU(2)$ symmetry, the sine-Gordon model (\ref{LSG}) is related via bosonization to the 
$\grSU(2)$ Thirring model \cite{Amit:1979ab,Banks:1975xs}, which contains two massless Dirac fermions $\psi^a$. Their interaction is $\sim\sum_{i=1}^3 J^i J^i$ where the
$\grSU(2)$ currents are $J^i= \frac{1}{2} \bar \psi^a \sigma^i_{ab} \psi^b$.

The $\beta$-function for the running mass parameter $\overline m$ can be inferred from \eqref{Gotalphadelta} and \eqref{Beta}:
\begin{align}\label{eq:rg}
\beta_{\overline m} =  M \frac{d\overline m}{d M} = -\frac{e^\gamma I_s}{4g^2}\overline m^3 \,, 
\end{align}
where $M$ is the RG scale. Thus, the interaction strength in the effective sine-Gordon model, and equivalently in the $\grSU(2)$ Thirring model, is asymptotically free.  The interaction strength formally diverges far in the IR, at the scale comparable to the mass gap (this scale is analogous to $\Lambda_{\rm QCD}$):
\begin{align}\label{eq:smallgap}
	E_{\rm gap}\sim e^{-A \frac{g^2}{m^2}} \,,  \qquad A= \frac{2e^{-\gamma}}{I_s}\approx0.111 \,.
\end{align}
This exponentially small mass gap implies that, for small $ m$, the correlation length diverges as $\xi \sim \frac{1}{E_\text{gap}} \sim  e^{A \frac{g^2}{m^2}}$.  Similarly, at small $m$ all observables can be expressed in terms of the energy scale $E_\text{gap}$.  For instance, since $\overline\Psi_\alpha\Psi_\alpha$ flows to an operator of dimension $\Delta_+ = 2$ in the $c=1$ theory at $m=0$, we must have $\langle \overline\Psi_\alpha\Psi_\alpha \rangle \sim E_\text{gap}^{\Delta_+} \sim e^{-2 A \frac{g^2}{m^2}}$.  Likewise, the operator $\Psi_1 \Psi_1 - \Psi_2 \Psi_2$ that takes us away from the $m_1 = m_2$ line in Figure~\eqref{PDFigure} flows to an operator of dimension $\Delta_- = 1/2$.  This allows us to estimate that the width of the symmetry breaking region is $\Delta m \sim E_\text{gap}^{2 - \Delta_-} = e^{- \frac{3A}{2} \frac{g^2}{m^2}}$.

\begin{figure}
	\centering
	\includegraphics[width=\linewidth]{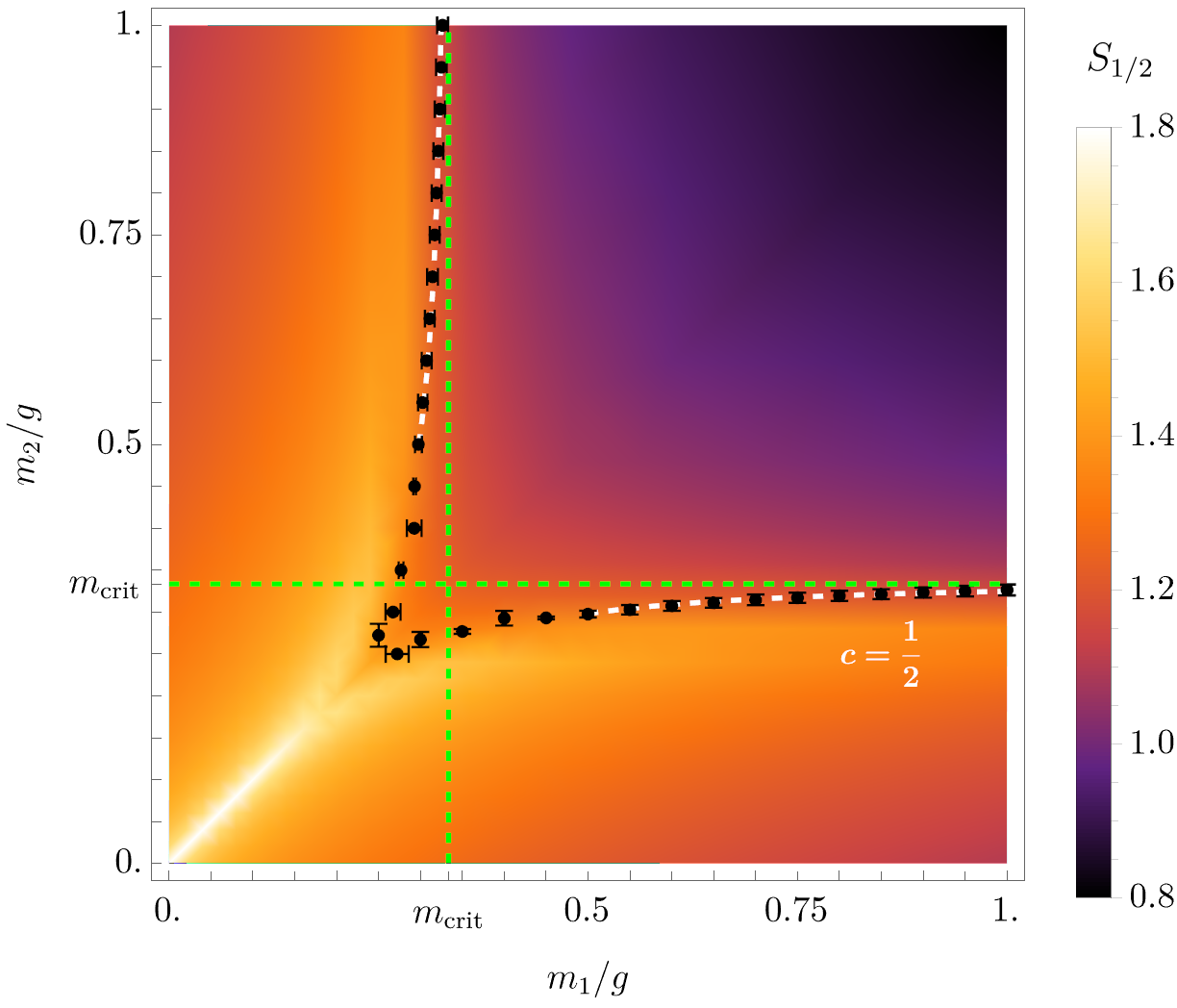}
	\caption{The heat map depicts the entanglement entropy $S_{1/2}(N=216, a = 0.3)$ with open boundary conditions as a function of the fermion masses at $\theta = \pi$. The black points are estimates of the location of the $c = \frac{1}{2}$ critical curve in the continuum limit $a\to 0$.  
The asymptotic shape of the curves agrees with \eqref{m1crit}.
For $m\ll g$, the two $c = \frac{1}{2}$ critical curves become exponentially close to each other.
}
	\label{fig:critical_curve}
\end{figure}

That charge conjugation symmetry is spontaneously broken for any $m>0$ can can be seen from \eqref{eq:sine-gordon}.  Indeed, in the bosonized description at $\theta = \pi$, the charge conjugation symmetry $C$ acts as
 \es{CThetaPi}{
   A_\mu \to - A_\mu \,, \quad \phi_+ \to - \phi_+ \,, \quad
    \phi_- \to -\phi_- + \sqrt{\frac{\pi}{2}}  \,.
 }
This is clearly a symmetry of the Lagrangian \eqref{BosLag} and also of the effective Lagrangian \eqref{eq:sine-gordon}.  However, over the range of one period $\phi_- \in [0, \sqrt{2 \pi}]$, the potential $\sim -\cos(\sqrt{8 \pi} \phi_-)$  in \eqref{eq:sine-gordon} has two minima, one at $\phi_- = 0$ and one $\phi_- =  \sqrt{\frac{\pi}{2}}$.  These minima are exchanged by the symmetry $C$ in \eqref{CThetaPi}.  Semi-classically, we thus have two vacua in which $C$ is broken spontaneously.

The spontaneous breaking of $C$ in the two-flavor Schwinger model provides a nice analogy to the breaking of $CP$ and presence of two degenerate vacua 
in 4D QCD with $\theta=\pi$ and two light flavors \cite{Creutz:1995wf}.  The height of the barrier separating the two symmetry breaking vacua is of order $m^2$, just as in QCD \cite{Creutz:1995wf,Smilga:1998dh}.
The $CP$ violation \cite{Dashen:1970et} can be seen using the chiral Lagrangian for QCD, and the zero-temperature phase diagram as a function of light quark masses $m_u$ and $m_d$ has a similar structure \cite{Creutz:2010ts,Creutz:2018vgl}
to our Figure \ref{PDFigure}. The boundaries of the region where $CP$ is spontaneosuly broken can be found from the condition that the mass of the neutral pion vanishes there. The width of the symmetry broken region is found to behave as $(m_u+m_d)^2/f_\pi $, which is parametrically much bigger than the exponentially small width that we find in the Schwinger model.  

\section{Numerical results}
\label{NUMERICS}

We study the $N_f = 2$ Schwinger model numerically using the lattice Hamiltonian \eqref{eq:gauge_hamiltonian} (see also \cite{Funcke:2023lli}). 
\begin{figure}[t]
	\centering
	\includegraphics[width=0.8\linewidth]{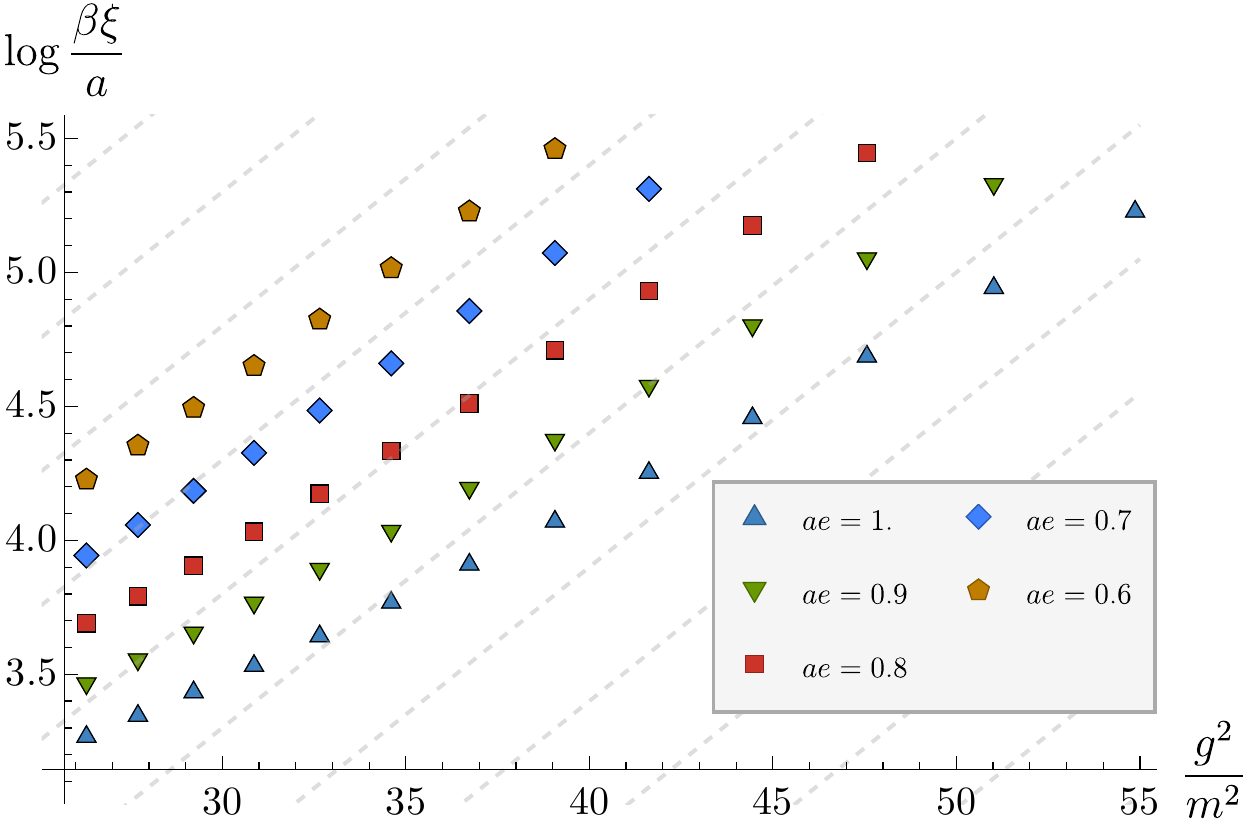}
	\caption{The estimated correlation length (up to a constant factor $\beta$) for $m_1 = m_2 \equiv m$ as a function of $m^{-2}$, showing a scaling of the form $\xi \sim e^{A \frac{g^2}{m^2}}$, as inferred from \eqref{eq:smallgap}. The dashed lines have slope $A \approx 0.111$. Extrapolating the slope from the lattice calculation to $a\to 0$ gives 0.11(1), in agreement with the theoretical value.}
	\label{fig:corr_length}
\end{figure}
While the one-flavor model can be studied efficiently via exact diagonalization \cite{Hamer:1982mx,Dempsey:2022nys}, with two flavors the number of states grows so quickly with the number $N$ of lattice sites that this becomes impractical. Instead, we employ tensor network methods, using a matrix product state (MPS) ansatz to approximate the ground state \cite{Banuls:2013jaa,Funcke:2023lli}. To optimize the MPS ansatz, we use \texttt{ITensors.jl} \cite{ITensor-r0.3,ITensor}. We use open boundary conditions, since this allows us to study the behavior of much larger lattices.

The MPS form of the ground state makes it especially simple to calculate the entanglement entropy for a left-right bipartition of the open chain. Let $S_x(N, a)$ denote the entanglement entropy for a subsystem of the leftmost $xN$ sites in a chain of $N$ sites with lattice spacing $a$. Then, at a critical point with central charge $c$, the entropy is expected to grow like \cite{Calabrese:2009qy}
\es{eq:critical_entanglement}{
	S_x(N, a) = \frac{c}{6}\log\left(\frac{2N}{\pi}\sin \pi x\right) + \text{const}.
}
At any other point, this logarithmic growth of the entropy will plateau when $N \sim \frac{\xi}{a}$, where $\xi$ is the correlation length. By combining this result with a finite-size scaling analysis, one can derive very precise estimates for the locations of critical points in the continuum theory from values of the entanglement entropy on a finite lattice \cite{Campostrini:2014lta}.

In Figure~\ref{fig:critical_curve}, we show the behavior of the entanglement entropy for a fixed lattice, along with the precise estimate of the critical curve obtained via the intersection method outlined in \cite{Campostrini:2014lta}.  The finite-size scaling analysis confirms that this curve has $c = \frac{1}{2}$. 
By fitting the leading large-mass behavior of this curve, we find
\es{eq:critical_curve_estimate}{
	m_{2,\text{cr}}(m_1) = 0.335(4) - 0.0097(17)\left(\frac{g}{m_1}\right)^2,
}
and the coefficient of $g^2/m_1^{2}$ is in good agreement with the value $\frac{0.3335}{12\pi} \approx 0.0088$ predicted from \eqref{m1crit}. 

We can also use lattice calculations of the entanglement entropy to estimate the growth of the correlation length for small $m_1 = m_2=m$. For a fixed lattice, we can compare the dependence of the entanglement entropy on the subsystem size with \eqref{eq:critical_entanglement} to obtain an estimate $c_\text{est}$ for the central charge. Anywhere away from a critical point, this estimate will tend to zero around $N\sim \frac{\xi}{a}$. We can thus take a fiducial cutoff for $c_\text{est}$, and define $\frac{\beta\xi}{a}$ as the lattice size when $c_\text{est}$ crosses below this cutoff, where $\beta$ is an unknown constant.

Figure \ref{fig:corr_length} shows this estimate of the logarithm of the correlation length along the $\grSU(2)$-invariant line as a function of $m^{-2}$. The linear behavior suggests a scaling of the form $\xi \sim e^{A \frac{g^2}{m^2}}$ at small $m/g$, as explained after \eqref{eq:smallgap}. Furthermore, extrapolating the slope as $m\to 0$ to the continuum limit $a \to 0$ gives $A =0.11(1)$, in good agreement with the theoretical value in \eqref{eq:smallgap}.

\section{Discussion}

In this paper, we presented analytical and numerical evidence for the phase diagram of the $N_f = 2$ Schwinger model at $\theta = \pi$ shown in Figure~\ref{PDFigure}.  The behavior we find is quite different from that at $\theta\neq \pi$: along the $\grSU(2)$-invariant line the theory contains a nearly marginal operator which leads to logarithmic RG flow of BKT type.  As a result, for $m\ll g$ the mass gap is exponentially small, $\sim e^{- A g^2/m^2}$.  Along this $\grSU(2)$-symmetric line, Georgi  \cite{Georgi:2022sdu} calculated the anomalous dimensions of operators perturbatively in powers of $(m/g)^2$.   The fact that the mass gap is exponentially small makes the theory for $m\ll g$ ``nearly conformal" in a large range of energies, so that perturbative anomalous dimension calculations should be parametrically reliable.   We thus hope that the calculations of \cite{Georgi:2022sdu} can be checked numerically using the lattice Hamiltonian setup, but we leave this question for future work.

We find that the $\mathbb{Z}_2$ charge conjugation symmetry is spontaneously broken in the entire shaded region of the phase diagram in Figure~\ref{PDFigure}.  This region becomes exponentially narrow near $m=0$ and is bounded by 2D Ising CFTs.    It is interesting to ask how the addition to the action of 4-fermion operators may change this phase diagram. We also leave this question for future work.

\section*{Acknowledgments}

We are grateful to Michael Creutz, Howard Georgi, Etsuko Itou, Andrei Katsevich, Zohar Komargodski, Lenny Susskind, Yuya Tanizaki, Grisha Tarnopolsky, and Edward Witten for very useful discussions. We thank a referee for important comments and pointing out very useful references.
This work was supported in part by the US National Science Foundation under Grants No.~PHY-2111977 and PHY-2209997, and by the Simons Foundation Grants 
No.~488653 and 917464.  RD was also supported in part by an NSF Graduate Research Fellowship.

\onecolumngrid
\vspace{1in}
\twocolumngrid

\bibliographystyle{ssg}
\bibliography{main}

\begingroup\raggedright\begin{thebibliography}{10}

\bibitem{Schwinger:1962tp}
J.~S. Schwinger, ``{Gauge Invariance and Mass. 2.},'' {\em Phys. Rev.} {\bf
  128} (1962) 2425--2429.

\bibitem{Lowenstein:1971fc}
J.~H. Lowenstein and J.~A. Swieca, ``{Quantum electrodynamics in
  two-dimensions},'' {\em Annals Phys.} {\bf 68} (1971) 172--195.

\bibitem{Casher:1974vf}
A.~Casher, J.~B. Kogut, and L.~Susskind, ``{Vacuum polarization and the absence
  of free quarks},'' {\em Phys. Rev. D} {\bf 10} (1974) 732--745.

\bibitem{Coleman:1975pw}
S.~R. Coleman, R.~Jackiw, and L.~Susskind, ``{Charge Shielding and Quark
  Confinement in the Massive Schwinger Model},'' {\em Annals Phys.} {\bf 93}
  (1975) 267.

\bibitem{Kogut:1974ag}
J.~B. Kogut and L.~Susskind, ``{Hamiltonian Formulation of Wilson's Lattice
  Gauge Theories},'' {\em Phys. Rev. D} {\bf 11} (1975) 395--408.

\bibitem{Banks:1975gq}
T.~Banks, L.~Susskind, and J.~B. Kogut, ``{Strong Coupling Calculations of
  Lattice Gauge Theories: (1+1)-Dimensional Exercises},'' {\em Phys. Rev. D}
  {\bf 13} (1976) 1043.

\bibitem{Banuls:2019bmf}
M.~C. Ba\~nuls {\em et.~al.}, ``{Simulating Lattice Gauge Theories within
  Quantum Technologies},'' {\em Eur. Phys. J. D} {\bf 74} (2020), no.~8 165,
  \href{https://arxiv.org/abs/1911.00003}{{\tt 1911.00003}}.

\bibitem{Coleman:1976uz}
S.~R. Coleman, ``More {About} the {Massive} {Schwinger} {Model},'' {\em Annals
  Phys.} {\bf 101} (1976) 239.

\bibitem{Gepner:1984au}
D.~Gepner, ``{Nonabelian Bosonization and Multiflavor {QED} and {QCD} in
  Two-dimensions},'' {\em Nucl. Phys. B} {\bf 252} (1985) 481--507.

\bibitem{Affleck:1985wa}
I.~Affleck, ``{On the Realization of Chiral Symmetry in (1+1)-dimensions},''
  {\em Nucl. Phys. B} {\bf 265} (1986) 448--468.

\bibitem{Georgi:2007ek}
H.~Georgi, ``{Unparticle physics},'' {\em Phys. Rev. Lett.} {\bf 98} (2007)
  221601, \href{https://arxiv.org/abs/hep-ph/0703260}{{\tt hep-ph/0703260}}.

\bibitem{Georgi:2019tch}
H.~Georgi and B.~Noether, ``{Non-perturbative Effects and Unparticle Physics in
  Generalized Schwinger Models},'' \href{https://arxiv.org/abs/1908.03279}{{\tt
  1908.03279}}.

\bibitem{Georgi:2020jik}
H.~Georgi, ``{Automatic Fine-Tuning in the Two-Flavor Schwinger Model},'' {\em
  Phys. Rev. Lett.} {\bf 125} (2020), no.~18 181601,
  \href{https://arxiv.org/abs/2007.15965}{{\tt 2007.15965}}.

\bibitem{Georgi:2022sdu}
H.~Georgi, ``{Mass perturbation theory in the 2-flavor Schwinger model with
  opposite masses with a review of the background},'' {\em JHEP} {\bf 10}
  (2022) 119, \href{https://arxiv.org/abs/2206.14691}{{\tt 2206.14691}}.

\bibitem{Steinhardt:1977tx}
P.~J. Steinhardt, ``{SU(2) Flavor Schwinger Model on the Lattice},'' {\em Phys.
  Rev. D} {\bf 16} (1977) 1782.

\bibitem{Smilga:1992hx}
A.~V. Smilga, ``On the fermion condensate in {Schwinger} model,'' {\em Phys.
  Lett. B} {\bf 278} (1992) 371--376.

\bibitem{Hetrick:1995wq}
J.~E. Hetrick, Y.~Hosotani, and S.~Iso, ``{The Massive multi - flavor Schwinger
  model},'' {\em Phys. Lett. B} {\bf 350} (1995) 92--102,
  \href{https://arxiv.org/abs/hep-th/9502113}{{\tt hep-th/9502113}}.

\bibitem{Smilga:1998dh}
A.~V. Smilga, ``{QCD at theta similar to pi},'' {\em Phys. Rev. D} {\bf 59}
  (1999) 114021, \href{https://arxiv.org/abs/hep-ph/9805214}{{\tt
  hep-ph/9805214}}.

\bibitem{Hosotani:1998kd}
Y.~Hosotani, ``{Antiferromagnetic S = 1/2 Heisenberg chain and the two flavor
  massless Schwinger model},'' {\em Phys. Rev. B} {\bf 60} (1999) 6198--6199,
  \href{https://arxiv.org/abs/hep-th/9809066}{{\tt hep-th/9809066}}.

\bibitem{Berruto:1999ga}
F.~Berruto, G.~Grignani, G.~W. Semenoff, and P.~Sodano, ``{On the
  correspondence between the strongly coupled two flavor lattice Schwinger
  model and the Heisenberg antiferromagnetic chain},'' {\em Annals Phys.} {\bf
  275} (1999) 254--296, \href{https://arxiv.org/abs/hep-th/9901142}{{\tt
  hep-th/9901142}}.

\bibitem{Hip:2021jgp}
I.~Hip, J.~F.~N. Castellanos, and W.~Bietenholz, ``{Finite temperature and
  $\delta$-regime in the 2-flavor Schwinger model},'' {\em PoS} {\bf
  LATTICE2021} (2022) 279, \href{https://arxiv.org/abs/2109.13468}{{\tt
  2109.13468}}.

\bibitem{Albergo:2022qfi}
M.~S. Albergo, D.~Boyda, K.~Cranmer, D.~C. Hackett, G.~Kanwar, S.~Racani\`ere,
  D.~J. Rezende, F.~Romero-L\'opez, P.~E. Shanahan, and J.~M. Urban,
  ``{Flow-based sampling in the lattice Schwinger model at criticality},''
  \href{https://arxiv.org/abs/2202.11712}{{\tt 2202.11712}}.

\bibitem{Funcke:2023lli}
L.~Funcke, K.~Jansen, and S.~K\"uhn, ``{Exploring the CP-Violating Dashen Phase
  in the Schwinger Model with Tensor Networks},''
  \href{https://arxiv.org/abs/2303.03799}{{\tt 2303.03799}}.

\bibitem{Dashen:1970et}
R.~F. Dashen, ``{Some features of chiral symmetry breaking},'' {\em Phys. Rev.
  D} {\bf 3} (1971) 1879--1889.

\bibitem{Creutz:1995wf}
M.~Creutz, ``{Quark masses and chiral symmetry},'' {\em Phys. Rev. D} {\bf 52}
  (1995) 2951--2959, \href{https://arxiv.org/abs/hep-th/9505112}{{\tt
  hep-th/9505112}}.

\bibitem{Creutz:2010ts}
M.~Creutz, ``{Quark mass dependence of two-flavor QCD},'' {\em Phys. Rev. D}
  {\bf 83} (2011) 016005, \href{https://arxiv.org/abs/1010.4467}{{\tt
  1010.4467}}.

\bibitem{Creutz:2018vgl}
M.~Creutz, ``{CP violation in QCD},'' {\em PoS} {\bf Confinement2018} (2018)
  171, \href{https://arxiv.org/abs/1810.03543}{{\tt 1810.03543}}.

\bibitem{Gaiotto:2017yup}
D.~Gaiotto, A.~Kapustin, Z.~Komargodski, and N.~Seiberg, ``{Theta, Time
  Reversal, and Temperature},'' {\em JHEP} {\bf 05} (2017) 091,
  \href{https://arxiv.org/abs/1703.00501}{{\tt 1703.00501}}.

\bibitem{Komargodski:2017dmc}
Z.~Komargodski, A.~Sharon, R.~Thorngren, and X.~Zhou, ``{Comments on Abelian
  Higgs Models and Persistent Order},'' {\em SciPost Phys.} {\bf 6} (2019),
  no.~1 003, \href{https://arxiv.org/abs/1705.04786}{{\tt 1705.04786}}.

\bibitem{Sulejmanpasic:2020lyq}
T.~Sulejmanpasic, D.~G\"oschl, and C.~Gattringer, ``{First-Principles
  Simulations of 1+1D Quantum Field Theories at $\theta=\pi$ and Spin
  Chains},'' {\em Phys. Rev. Lett.} {\bf 125} (2020), no.~20 201602,
  \href{https://arxiv.org/abs/2007.06323}{{\tt 2007.06323}}.

\bibitem{Dempsey:2022nys}
R.~Dempsey, I.~R. Klebanov, S.~S. Pufu, and B.~Zan, ``{Discrete chiral symmetry
  and mass shift in the lattice Hamiltonian approach to the Schwinger model},''
  {\em Phys. Rev. Res.} {\bf 4} (2022), no.~4 043133,
  \href{https://arxiv.org/abs/2206.05308}{{\tt 2206.05308}}.

\bibitem{Byrnes:2002gj}
T.~Byrnes, P.~Sriganesh, R.~J. Bursill, and C.~J. Hamer, ``{Density matrix
  renormalization group approach to the massive Schwinger model},'' {\em Nucl.
  Phys. B Proc. Suppl.} {\bf 109} (2002) 202--206,
  \href{https://arxiv.org/abs/hep-lat/0201007}{{\tt hep-lat/0201007}}.

\bibitem{Byrnes:2002nv}
T.~Byrnes, P.~Sriganesh, R.~J. Bursill, and C.~J. Hamer, ``{Density matrix
  renormalization group approach to the massive Schwinger model},'' {\em Phys.
  Rev. D} {\bf 66} (2002) 013002,
  \href{https://arxiv.org/abs/hep-lat/0202014}{{\tt hep-lat/0202014}}.

\bibitem{Peskin:1995ev}
M.~E. Peskin and D.~V. Schroeder, {\em {An Introduction to quantum field
  theory}}.
\newblock Addison-Wesley, Reading, USA, 1995.

\bibitem{Cheng:2022sgb}
M.~Cheng and N.~Seiberg, ``{Lieb-Schultz-Mattis, Luttinger, and 't Hooft --
  anomaly matching in lattice systems},''
  \href{https://arxiv.org/abs/2211.12543}{{\tt 2211.12543}}.

\bibitem{Kosterlitz:1974sm}
J.~M. Kosterlitz, ``{The Critical properties of the two-dimensional x y
  model},'' {\em J. Phys. C} {\bf 7} (1974) 1046--1060.

\bibitem{Amit:1979ab}
D.~J. Amit, Y.~Y. Goldschmidt, and G.~Grinstein, ``{Renormalization Group
  Analysis of the Phase Transition in the 2D Coulomb Gas, Sine-Gordon Theory
  and xy Model},'' {\em J. Phys. A} {\bf 13} (1980) 585.

\bibitem{Banks:1975xs}
T.~Banks, D.~Horn, and H.~Neuberger, ``{Bosonization of the SU(N) Thirring
  Models},'' {\em Nucl. Phys. B} {\bf 108} (1976) 119.

\bibitem{Hamer:1982mx}
C.~J. Hamer, J.~B. Kogut, D.~P. Crewther, and M.~M. Mazzolini, ``{The Massive
  Schwinger Model on a Lattice: Background Field, Chiral Symmetry and the
  String Tension},'' {\em Nucl. Phys. B} {\bf 208} (1982) 413--438.

\bibitem{Banuls:2013jaa}
M.~C. Ba\~nuls, K.~Cichy, K.~Jansen, and J.~I. Cirac, ``{The mass spectrum of
  the Schwinger model with Matrix Product States},'' {\em JHEP} {\bf 11} (2013)
  158, \href{https://arxiv.org/abs/1305.3765}{{\tt 1305.3765}}.

\bibitem{ITensor-r0.3}
M.~Fishman, S.~R. White, and E.~M. Stoudenmire, ``Codebase release 0.3 for
  {ITensor},'' {\em SciPost Phys. Codebases} (2022) 4--r0.3. Publisher:
  SciPost.

\bibitem{ITensor}
M.~Fishman, S.~R. White, and E.~M. Stoudenmire, ``The {ITensor} software
  library for tensor network calculations,'' {\em SciPost Phys. Codebases}
  (2022) 4. Publisher: SciPost.

\bibitem{Calabrese:2009qy}
P.~Calabrese and J.~Cardy, ``{Entanglement entropy and conformal field
  theory},'' {\em J. Phys. A} {\bf 42} (2009) 504005,
  \href{https://arxiv.org/abs/0905.4013}{{\tt 0905.4013}}.

\bibitem{Campostrini:2014lta}
M.~Campostrini, A.~Pelissetto, and E.~Vicari, ``{Finite-size scaling at quantum
  transitions},'' {\em Phys. Rev. B} {\bf 89} (2014), no.~9 094516,
  \href{https://arxiv.org/abs/1401.0788}{{\tt 1401.0788}}.

\end{thebibliography}\endgroup

\end{document}